

\def\bra#1{{\langle#1\vert}}
\def\ket#1{{\vert#1\rangle}}
\def\coeff#1#2{{\scriptstyle{#1\over #2}}}

\def\sst#1{{\scriptscriptstyle #1}}

\def\nubar{{\bar\nu}}

\def\alr{{A_\sst{LR}}}

\def\evec{{\vec e}}

\def\mn{{m_\sst{N}}}
\def\mns{{m^2_\sst{N}}}

\def\ubar{{\bar u}}
\def\dbar{{\bar d}}
\def\sbar{{\bar s}}
\def\qbar{{\bar q}}

\def\sqr#1#2{{\vcenter{\vbox{\hrule height.#2pt
		\hbox{\vrule width.#2pt height#1pt \kern#1pt
			\vrule width.#2pt}
		\hrule height.#2pt}}}}

\def\sst#1{{\scriptscriptstyle #1}}

\def\mpi{{m_\pi}}

\def\mn{{m_\sst{N}}}
\def\mns{{m^2_\sst{N}}}

\def\gpnn{{g_{\sst{NN}\pi}}}

\def\sst#1{{\scriptscriptstyle #1}}

\def\mn{{m_\sst{N}}}
\def\mns{{m_\sst{N}^2}}

\def\mpi{{m_\pi}}

\def\gpnn{{g_{\pi\sst{NN}}}}

\def\ubar{{\bar u}}

\def\GAS{{G_\sst{A}^{(s)}}}

\def\mustr{{\mu_s}}

\def\rhostr{{\rho_s}}

\def\GEn{{G_\sst{E}^n}}

\def\bra#1{{\langle#1\vert}}
\def\ket#1{{\vert#1\rangle}}
\def\coeff#1#2{{\scriptstyle{#1\over #2}}}

\def\sst#1{{\scriptscriptstyle #1}}

\def\nubar{{\bar\nu}}

\def\alr{{A_\sst{LR}}}

\def\evec{{\vec e}}

\def\mn{{m_\sst{N}}}
\def\mns{{m^2_\sst{N}}}

\def\ubar{{\bar u}}
\def\dbar{{\bar d}}
\def\sbar{{\bar s}}
\def\qbar{{\bar q}}

\def\sqr#1#2{{\vcenter{\vbox{\hrule height.#2pt
		\hbox{\vrule width.#2pt height#1pt \kern#1pt
			\vrule width.#2pt}
		\hrule height.#2pt}}}}

\def\sst#1{{\scriptscriptstyle #1}}

\def\mpi{{m_\pi}}

\def\mn{{m_\sst{N}}}
\def\mns{{m^2_\sst{N}}}

\def\gpnn{{g_{\sst{NN}\pi}}}

\def\sst#1{{\scriptscriptstyle #1}}

\def\mn{{m_\sst{N}}}
\def\mns{{m_\sst{N}^2}}

\def\mpi{{m_\pi}}

\def\gpnn{{g_{\pi\sst{NN}}}}

\def\ubar{{\bar u}}

\def\GAS{{G_\sst{A}^{(s)}}}

\def\mustr{{\mu_s}}

\def\rhostr{{\rho_s}}

\def\GEn{{G_\sst{E}^n}}


\def\evec{{\vec e}}

\def\mustr{{\mu_s}}

\def\nubar{{\bar\nu}}

\def\mbar{{\overline m}}
\def\lambar{{\overline\Lambda}}

\def\NPB#1{{\it Nucl. Phys.} {\bf B#1} }

\def\ZPC#1{{\it Z. f\"ur Phys.} {\bf C#1} }

\def\qbar{{\bar q}}

\hfuzz=50pt

\vsize=7.5in
\hsize=5.6in
\magnification=1200
\tolerance=10000

\baselineskip 12pt plus 1pt minus 1pt
\pageno=0
\centerline{\bf Stranger Still: Kaon Loops and Strange Quark}
\smallskip
\centerline{{\bf Matrix Elements of the Nucleon}\footnote{*}{This
work is supported in part by funds
provided by the U. S. Department of Energy (D.O.E.) under contracts
\#DE-AC02-76ER03069 and \#DE-AC05-84ER40150.}}
\vskip 24pt
\centerline{M. J. Musolf}
\vskip 12pt
\centerline{\it Department of Physics}
\centerline{\it Old Dominion University}
\centerline{\it Norfolk, VA \ \   23529\ \ \ U.S.A.}
\centerline{\it and}
\centerline{\it CEBAF Theory Group, MS-12H}
\centerline{\it 12000 Jefferson Ave.}
\centerline{\it Newport News, VA \ \ 23606\ \ \ U.S.A.}
\vskip 12pt
\centerline{\it and}
\vskip 12pt
\centerline{M. Burkardt}
\vskip 12pt
\centerline{\it Center for Theoretical Physics}
\centerline{\it Laboratory for Nuclear Science}
\centerline{\it and Department of Physics}
\centerline{\it Massachusetts Institute of Technology}
\centerline{\it Cambridge, Massachusetts\ \ 02139\ \ \ U.S.A.}
\vskip 1.5in
\centerline{}
\vfill
\noindent CEBAF \#TH-93-01\hfill January, 1993
\eject
\baselineskip 16pt plus 2pt minus 2pt
\centerline{\bf ABSTRACT}
\medskip
Intrinsic strangeness contributions to low-energy strange quark matrix
elements of the nucleon are modelled using kaon loops and
meson-nucleon vertex functions taken from nucleon-nucleon and
nucleon-hyperon scattering.
A comparison with pion loop contributions to the
nucleon electromagnetic (EM) form factors indicates the presence of significant
SU(3)-breaking in the mean-square charge radii. As a numerical consequence,
the kaon loop contribution to the mean square Dirac strangeness radius is
significantly smaller than could be observed with parity-violating elastic
$\evec p$ experiments planned for CEBAF, while the contribution to the Sachs
radius is large enough to be observed with PV electron scattering from
$(0^+,0)$ nuclei. Kaon loops generate a strange magnetic moment of the same
scale as the isoscalar EM magnetic moment and a strange axial vector
form factor having roughly one-third
magnitude extracted from $\nu p/\nubar p$ elastic scattering.
In the chiral limit, the loop contribution to the fraction of the nucleon's
scalar density arising from strange quarks has roughly the same magnitude as
the value extracted from analyses of $\Sigma_{\pi N}$. The importance of
satisfying the Ward-Takahashi Identity, not obeyed by previous calculations,
is also illustrated, and the sensitivity of results to input parameters is
analyzed.
\vfill
\eject

\noindent{\bf 1. Introduction}

	Extractions of the strange quark scalar density from $\Sigma_{\pi N}$
[1,2], the strange quark axial vector form
factor from elastic $\nu p/\nubar p$ cross section measurements [3], and the
strange-quark contribution to the proton spin $\Delta s$ from the
EMC measurement of the $g_1$ sum [4], suggest a need to account explicitly
for the presence of strange quarks in the nucleon in describing its low-
energy properties. These analyses have motivated suggestions for measuring
strange quark vector current matrix elements of the nucleon with semi-leptonic
neutral current  scattering [5]. The goal of the SAMPLE experiment
presently underway at MIT-Bates [6] is to constrain the strange quark magnetic
form factor at low-$|q^2|$, and three experiments have been planned and/or
proposed at CEBAF with the objective of constraining the nucleon's mean
square \lq\lq strangeness radius" [7-9]. In addition, a new determination
of the strange quark axial vector form factor at significantly lower-$|q^2|$
than was obtained from the experiment of Ref.~[3] is expected from LSND
experiment at LAMPF [10].

	At first glance, the existence of non-negligible low-energy strange
quark matrix elements of the nucleon is rather surprising, especially in
light of the success with which constituent quark models account for other
low-energy properties of the nucleon and its excited states. Theoretically,
one might attempt to understand the possibility of large strange matrix
elements from two perspectives associated, respectively, with the high-
and low-momentum components of a virtual $s\sbar$ pair in the nucleon.
Contributions from the high-momentum component may be viewed as
\lq\lq extrinsic" to the nucleon's wavefunction, since the lifetime of the
virtual pair at high-momentum scales is shorter than the interaction time
associated with the formation of hadronic states [11]. In an effective theory
framework, the extrinsic, high-momentum contributions renormalize operators
involving explicitly only light-quark degrees of freedom [5, 12]. At
low-momentum scales, a virtual pair lives a sufficiently long time to permit
formation strange hadronic components ({\it e.g.}, a $K\Lambda$ pair) of the
nucleon wavefunction [13]. While this division between \lq\lq extrinsic" and
\lq\lq intrinsic" strangeness is not rigorous, it does provide a qualitative
picture which suggests different approaches to estimating nucleon strange
matrix
elements.

	In this note we consider intrinsic strangeness contributions to
the matrix elements $\bra{N} \sbar\Gamma s\ket{N}$ ($\Gamma\ =\ 1$,
$\gamma_\mu$, $\gamma_\mu\gamma_5$) arising from kaon-strange baryon loops.
Our calculation is intended to complement pole [14] and Skyrme [15] model
estimates as well as to quantify the simple picture of nucleon strangeness
arising from a kaon cloud. Although loop estimates have been carried out
previously [16, 17], ours differs from others in several respects. First,
we assume that nucleon electromagnetic (EM) and
and weak neutral current (NC) form factors receive
contributions from a variety of sources ({\it e.g.}, loops and poles),
so we make no attempt to adjust the input parameters to reproduce known
form factors ({\it e.g.}, $\GEn$). Rather, we take our inputs from independent
sources, such as fits to baryon-baryon scattering and quark model
estimates where
needed. We compute pion loop contributions to the nucleon's EM form factors
using these input parameters and compare with the experimental values.
Such a comparison provides an indication of the extent to which loops account
for nucleon form factor physics generally and strangeness form factors in
particular. Second, we employ hadronic form factors at the meson nucleon
vertices and introduce \lq\lq seagull" terms in order to satisfy the
Ward-Takahashi (WT)
Identity in the vector current sector. Previous loop calculations employed
either a momentum cut-off in the loop integral [16] or meson-baryon form
factor [17] but did not satisfy the WT Identity. We find that failure to
satisfy the requirements of gauge invariance at this level
can significantly alter one's results. Finally, we include an estimate of
the strange quark scalar density which was not included in previous work.

\noindent{\bf 2. The calculation.}

The loop diagrams which we compute are shown in Fig. 1. In the case of
vector current matrix elements, all four diagrams contribute, including
the two seagull graphs (Fig. 1c,d) required by gauge invariance. For
axial vector matrix elements, only the loop of Fig. 1a contributes, since
$\bra{M} J_{\mu 5}\ket{M}\equiv 0$ for $M$ a pseudoscalar meson. The loops
1a and 1b contribute to $\bra{N}\sbar s\ket{N}$. In a world of point
hadrons satisfying SU(3) symmetry, the coupling of the lowest baryon and
meson octets is given by
$$
i{\cal L}_\sst{BBM}=D\ {\rm Tr}[(B{\bar B}+{\bar B}B)M]+
F\ {\rm Tr}[(B{\bar B}-{\bar B}B)M]\eqno(1)
$$
where $\sqrt{2}B=\sum_a \psi_a\lambda_a$ and $\sqrt{2}M=\sum_a\phi_a
\lambda_a$ give the octet of baryon and meson fields, respectively, and where
$D+F=\sqrt{2}\gpnn=19.025$ and $D/F=1.5$
according to Ref.~[18]. Under
this parameterization, one has $g_\sst{N\Sigma K}/g_\sst{N\Lambda K} = \sqrt{3}
(F-D)/(D+3F)\approx -1/5$, so that loops having an intermediate $K\Sigma$
state ought to be generically suppressed by a factor of $\sim 25$ with
respect to $K\Lambda$ loops. Analyses of $K+N$ \lq\lq strangeness exchange"
reactions, however, suggest a
serious violation of this SU(3) prediction [19],
and imply that neglect of $K\Sigma$ loops is not necessarily justified.
Nevertheless, we consider only $K\Lambda$ loops since we are interested
primarily in arriving at the order of magnitude and qualitative features of
loop contributions and not definitive numerical predictions.

	With point hadron vertices, power counting implies that loop
contributions to the mean square charge radius and magnetic moment are U.V.
finite. In fact, the pion loop contributions to the nucleon's EM charge
radius and magnetic moment have been computed previously in the limit of
point hadron vertices [20]. Loop contributions to axial vector and scalar
density matrix
elements, however, are U.V. divergent, necessitating use of
a cut-off procedure. To this end, we employ form factors at the meson-nucleon
vertices used in determination of the Bonn potential from $BB'$ scattering
($B$ and $B'$ are members of the lowest-lying baryon octet) [18]:
$$
g_\sst{NNM}\gamma_5\longrightarrow g_\sst{NNM}F(k^2)\gamma_5\ \ \ ,\eqno(2)
$$
where
$$
F(k^2)=\Bigl[{m^2-\Lambda^2\over k^2-\Lambda^2}\Bigr]\ \ \ ,\eqno(3)
$$
with $m$ and $k$ being the mass and
momentum, respectively, of the meson. The Bonn values for cut-off $\Lambda$
are typically in the range of 1 to 2 GeV. We note that this form
reproduces the point hadron coupling for on-shell mesons ($F(m^2)=1$).
An artifact of this choice is the vanishing of the form factors
(and all loop amplitudes) for $\Lambda=m$.
Consequently, when analyzing the $\Lambda$-dependence of our results below,
we exclude the region about $\Lambda=m$ as unphysical.

	For $\Lambda\to\infty$ (point hadrons), the total contribution from
diagrams 1a and 1b to vector current matrix elements satisfies the WT Identity
$q^\mu\Lambda(p,p')_\mu=Q[\Sigma(p')-\Sigma(p)]$, where $Q$ is the nucleon
charge associated with the corresponding vector current (EM, strangeness,
baryon number, {\it etc}).
For finite $\Lambda$, however, this identity is not satisfied
by diagrams 1a$+$1b alone; inclusion of seagull diagrams (1c,d) is required
in order to preserve it.
To arrive at the appropriate seagull vertices, we treat the momentum-space
meson-nucleon vertex functions as arriving from a phenomenological lagrangian
$$
i{\cal L}_\sst{BBM}\longrightarrow g_\sst{BBM}{\bar\psi}\gamma_5\psi
F(-\partial^2)\phi\ \ \ ,\eqno(4)
$$
where $\psi$ and $\phi$ are baryon and meson fields, respectively. The gauge
invariance of this lagrangian can be maintained via minimal substitution.
We replace the derivatives in the d'Alembertian by covariant derivatives,
expand $F(-D^2)$ in a power series, identify the terms linear in
the gauge field, express the resulting series in closed form, and
convert back to momentum space. With our choice for $F(k^2)$, this prodecure
leads to the seagull vertex
$$
-ig_\sst{BBM} Q F(k^2)\Bigl[{(q\pm 2k)_\mu\over (q\pm k)^2-\Lambda^2}\Bigr]
\eqno(5)
$$
where $q$ is the momentum of the gauge boson and where the upper (lower)
sign corresponds to an incoming (outgoing) meson of charge $Q$
(details of this procedure are given in Ref.~[21]).
With these vertices in diagrams 1c,d, the WT Identity
in the presence of meson nucleon form factors is restored. We note that
this prescription for satisfying gauge invariance is not unique;
the specific underlying dynamics which give rise to $F(k^2)$
could generate additional seagull terms whose contributions independently
satisfy the
WT Identity. Indeed, different models of hadron structure may lead to
meson-baryon form factors having a different form than our choice.
However, for the purposes of our calculation, whose spirit is to arrive at
order
of magnitude esitmates and qualitative features, the use of the Bonn form
factor plus minimal substitution is sufficient.

The strange vector, axial vector, and scalar density couplings to the
intermediate hardrons can be obtained with varying degrees of model-dependence.
Since we are interested only in the leading $q^2$-behavior of the nucleon
matrix elements as generated by the loops, we emply point couplings to the
intermediate meson and baryon. For the vector currents, one has
$\bra{\Lambda(p')}\sbar\gamma_\mu s\ket{\Lambda(p)} = f_\sst{V}{\bar U}(p')
\gamma_\mu U(p)$ and $\bra{K^0(p')}\sbar\gamma_\mu s
\ket{K^0(p)}={\tilde f_\sst{V}}(p+p')_\mu$ with $f_\sst{V}=-{\tilde f_\sst{V}}
=1$ in a convention where the $s$-quark has strangeness charge $+1$. The
vector couplings are determined simply by the net strangeness of the hadron,
independent of the details of any hadron model.

In the case of the axial vector, only the baryon coupling is required since
pseudoscalar mesons have no diagonal axial vector matrix elements. Our approach
in this case is to use a quark model to relate the \lq\lq bare" strange axial
vector coupling to the $\Lambda$ to the bare isovector axial vector matrix
element of the nucleon, where by \lq\lq bare" we mean that
the effect of meson loops has not been included. We then compute the loop
contributions to the ratio
$$
\eta_s = {\GAS(0)\over g_\sst{A}}\ \ \ ,\eqno(6)
$$
where $\GAS(q^2)$ is the strange quark axial vector form factor
(see Eq.~(10) below) and $g_\sst{A}=1.262$  [22]
is the proton's isovector axial vector form factor
at zero momentum transfer. Writing $\bra{\Lambda(p')}\sbar
\gamma_\mu\gamma_5 s\ket{\Lambda(p)}=f_\sst{A}^0{\bar U}(p')\gamma_\mu\gamma_5
U(p)$, one has in the quark model
$f_\sst{A}^0 = \int d^3x (u^2-\coeff{1}{3}\ell^2)$,
where $u$ ($\ell$) are the upper (lower) components of a quark in its
lowest energy configuration [23, 24, 25]. The quark model also predicts that
$g_\sst{A}^0 \equiv \coeff{5}{3}\int d^3x (u^2-\coeff{1}{3}\ell^2)$.
In the present calculation, we take the baryon octet to be SU(3) symmetric
({\it e.g.}, $\mn=m_\Lambda$), so that the quark wavefunctions
are the same for the nucleon and $\Lambda$.\footnote{\dag}{We investigate
the consequences of SU(3)-breaking in the baryon octet in a forthcoming
publication [21].}\ In this case,
one has $f_\sst{A}^0=\coeff{3}{5} g_\sst{A}^0$. This relation
holds in both the relativistic quark model and the simplest non-relativistic
quark model in which one simply drops the lower component contributions to
the quark model integrals. We will make the further assumption that
pseudoscalar meson loops generate the dominant renormalization of the bare
axial couplings.
The $\Lambda$ has no isovector axial vector matrix element, while
loops involving $K\Sigma$ intermediate states are suppressed in the SU(3)
limit as noted earlier. Under this assumption, then, only the $\pi N$
loop renormalizes the bare coupling, so that
$g_\sst{A}=g_\sst{A}^0[1+\Delta_\sst{A}^\pi(\Lambda)]$,
where $\Delta_\sst{A}^\pi(\Lambda)$ gives the contribution from the $\pi N$
loop with the bare coupling to the intermediate nucleon scaled out. In this
case, the ratio $\eta_s$ is essentially independent of the
actual numerical predictions for $f_\sst{A}^0$ and $g_\sst{A}^0$ in a given
quark model; only the spin-flavor factor $\coeff{3}{5}$ which relates the
two enters.

For the scalar density, we require point couplings to both the intermediate
baryon and meson. We write these couplings as $\bra{B(p')}\qbar q\ket{B(p)}
=f_\sst{S}^0 {\bar U(p')} U(p)$ and $\bra{M(p')}\qbar q\ket{M(p)} =
\gamma_\sst{M}$, where $B$ and $M$ denote the meson and baryon, respectively.
Our choices for $f_s^0$ and $\gamma_\sst{M}$ carry the most hadron
model-dependence of all our input couplings. To reduce the impact of this
model-dependence on our result, we again compute loop contributions to a
ratio, namely,
$$
R_s\equiv{\bra{N}\sbar s\ket{N}\over\bra{N} \ubar u +\dbar d+\sbar s
\ket{N}}\ \ \ .\eqno(7)
$$
In the language of Ref.~[2], on has $R_s=y/(2+y)$. Our aim in the present
work is to compute $R_s$ in a manner as free as possible from the
assumptions made in extracting this quantity from standard $\Sigma$-term
analyses. We therefore use the quark model to compute $f_\sst{S}^0$ and
$\gamma_\sst{M}$ rather than obtaining these parameters from a chiral
SU(3) fit to hadron mass splittings. We explore this alternative procedure,
along with the effects of SU(3)-breaking in the baryon octet, elsewhere [21].

In the limit of good SU(3) symmetry for the baryon octet,
the bare $\sbar s$ matrix element of the $\Lambda$ is
$f_\sst{S}^0 = \int d^3x (u^2-\ell^2)$.
Using the wavefunction normalization condition $\int d^3x (u^2 +
\ell^2) = 1$, together with the quark model expression for
$g_\sst{A}^0$, leads to
$$
f_\sst{S}^0 = \coeff{1}{2}
(\coeff{9}{5} g_\sst{A}^0 - 1)\ \ \ .\eqno(8)
$$
Neglecting loops, one has
$\bra{N}\ubar u +\dbar d+\sbar s\ket{N}=3f_\sst{S}^0$. We include loop
contributions to both the numerator and denominator of Eq.~(7).
Although the latter
turn out to be numerically unimportant, their inclusion guarantees that
$R_s$ is finite in the chiral limit.

Were the contribution from Fig. 1a dominant, the loop estimate of $R_s$
would be nearly independent of $f_\sst{S}^0$. The contribution from
Fig. 1b, however, turns out to have comparable magnitude. Consequently, we
are unable to minimize the hadron model-dependence in our estimate of $R_s$ to
the same extent we are able with $\eta_s$ and the vector current form
factors. To arrive at a value for $f_\sst{S}^0$, we consider three
alternatives: (A) Compute $f_\sst{S}^0$ using the MIT bag model value
for $g_\sst{A}^0$ and use a cut-off $\Lambda\sim\Lambda_{\rm Bonn}$ in
the meson-baryon form factor. This scenario
suffers from the conceptual ambiguity that the Bonn value for the
cut-off mass in $F(k^2)$ allows for virtual mesons of wavelength smaller
than the bag radius. (B) Compute $f_\sst{S}^0$ as in (A)
and take the cut-off $\Lambda\sim 1/R_{\rm bag}$. This approach follows
in the spirit of so-called \lq\lq chiral quark models", such as
that used in the calculation of Ref.~[17], which assume the virtual
pseudoscalar mesons are Goldstone bosons that live only outside the bag
and couple directly to the quarks at the bag surface. While
conceptually more satisfying than (A), this choice leads to a form
factor $F(k^2)$ inconsistent with $BB'$ scattering data. (C) First,
determine $g_\sst{A}^0$ assuming pion-loop dominance in the isovector
axial form factor, {\it i.e.}, $g_\sst{A}=g_\sst{A}^0[1+\Delta_\sst{A}^\pi
(\Lambda)]$. Second, use this value of $g_\sst{A}^0$ to determine
$f_\sst{S}^0$ via Eq.~(8).
Surprisingly, this procedure yields a value for
$g_\sst{A}^0$ very close to the MIT bag model value for $\Lambda\sim
\Lambda_{\rm Bonn}$ rather than $\Lambda\sim 1/R_{\rm bag}$ as one might
naively expect.

Since all three of these scenarios are consistent with the bag estimate
for $g_\sst{A}^0$ (and renormalization constant, $Z$ as discussed below)
we follow the \lq\lq improved bag" procedure of Ref.~[24] to obtain
$\gamma_\sst{M}$. The latter gives $\gamma_\sst{M}
\equiv\bra{M(p')}{\bar q} q\ket{M(p)} = 1.4/R$, where $R$ is the bag radius
for meson $M$. Using $R_\pi\approx R_\sst{K} \approx 3.5\ \hbox{GeV}^{-1}$,
one has $\gamma_\pi\approx\gamma_\sst{K}\approx 0.4 $ GeV. This procedure
involves a certain degree of theoretical uncertainty. The estimate for
$\gamma_\sst{K}$ ($\gamma_{\pi}$) is obtained by expanding the bag energy to
leading  non-trivial order in $m_\sst{K}$ and $m_s$ ($\mpi$ and $m_{u,d}$),
and we have at present no estimate of the corrections induced by higher-order
terms in these masses.

Using the above couplings, we compute the kaon loop contributions to the
strange quark scalar density as well as vector
and axial vector form factors. The latter are defined as
$$
\eqalignno{
\bra{N(p')}J_\mu(0)\ket{N(p)}&={\bar U}(p')\Bigl[F_1\gamma_\mu +{iF_2\over
	2\mn}\sigma_{\mu\nu}q^\nu\Bigr] U(p)& (9)\cr
\bra{N(p')}J_{\mu 5}(0)\ket{N(p)}&={\bar U}(p')\Bigl[G_\sst{A}\gamma_\mu
+G_\sst{P}{q_\mu\over\mn}\Bigr]\gamma_5 U(p)& (10)\cr}
$$
where $J_\mu$ is either the EM or strange quark vector current and
$J_{\mu 5}$ is the strange axial vector current. The induced pseudoscalar
form factor, $G_\sst{P}$, does not enter semi-leptonic neutral current
scattering processes at an observable level, so we do not discuss it here.
In the case of
the EM current, pion loop contributions to the neutron form factors
arise from the same set of diagrams as contribute to the strange vector
current matrix elements but with the replacements $K^0\to\pi^-$, $\Lambda
\to n$, and $g_\sst{N\Lambda K}\to \sqrt{2}\gpnn$. For the proton, one has
a $\pi^0$ in Fig. 1a and a $\pi^+$ in Figs. 1b-d. We quote results for
both Dirac and Pauli form factors as well as for the Sachs electric and
magnetic form factors [26], defined as $G_\sst{E}=F_1-\tau F_2$
and $G_\sst{M}=F_1+F_2$, respectively, where $\tau=-q^2/4\mns$ and
$q^2=\omega^2-|\vec q|^2$. We define the magnetic moment as
$\mu=G_\sst{M}(0)$ and dimensionless mean square Sachs and Dirac
charge radii (EM or strange) as
$$
\eqalignno{
\rho^{\rm sachs}& = {d G_\sst{E}(\tau)\over d\tau}\Bigr\vert_{\tau = 0}
&(11) \cr
\rho^{\rm dirac}& = {d F_1(\tau)\over d\tau}\Bigr\vert_{\tau = 0}\ \ \ .
&(12)\cr}
$$
The dimensionless radii are related to the dimensionfull mean square
radii by $< r^2 >_{\rm sachs} = 6\ d G_\sst{E}/dq^2= -(3/ 2\mns)\rho^{\rm
sachs}$
and similarly for the corresponding Dirac quantities. From these definitions,
one has $\rho^{\rm dirac}=\rho^{\rm sachs}+\mu$. To set the scales, we note
that the  Sachs EM charge radius of the neutron is $\rho_n^{\rm sachs}\approx
-\mu_n$, corresponding to an $< r^2_n >_{\rm sachs}$ of about -0.13
$\hbox{fm}^2$. Its Dirac EM charge radius, on the other hand, is nearly zero.
We note also that it is the Sachs, rather than the Dirac, mean square
radius which characterizes the spatial distribution of the corresponding
charge inside the nucleon, since it is the combination $F_1-\tau F_2$
which arises naturally in a non-relativisitc expansion of the time component
of Eq.~(9).

\noindent{\bf 3. Results and discussion}

	Our results are shown in Fig. 2, where we plot the different
strange matrix elements as a function of the form factor cut-off, $\Lambda$.
Although we quote results in Table I corresponding to the Bonn fit values for
$\Lambda$, we show the $\Lambda$-dependence away from $\Lambda_{\rm Bonn}$
to indicate the sensitivity to the cut-off. In each case, we plot two
sets of curves corresponding, respectively,
to $m=m_\sst{K}$ and $m=m_\pi$, in order to illustrate
the dependence on meson mass as well as to show the pion loop contributions
to the EM form factors. The dashed curves for the mean
square radius and magnetic moment give the values for $\Lambda\to\infty$,
corresponding to the point hadron calculation of Ref.~[20].
We reiterate that the zeroes arising at $\Lambda=
m$ are an unphysical artifact of our choice of nucleon-meson form factor,
and one should not draw conclusions from the curves in the vicinity of these
points.

In order to interpret our results, it is useful to refer to the analytic
expressions for the loops in various limits. The full analytic expressions
will appear in a forthcoming publication [21].
In the case of the vector current form factors, the
use of monopole meson-nucleon form factor plus minimal substitution leads to
the result that
$$
F_{(i)} = F_{(i)}^{\rm point}(m^2)-F_{(i)}^{\rm point}(\Lambda^2)+
(\Lambda^2-m^2){d\over d\Lambda^2}F_{(i)}^{\rm point}(\Lambda^2)\ \ \
,\eqno(13)
$$
where $F_{(i)}^{\rm point}(m^2)$ is the point hadron result of Ref.~[20].
It is straightforward to show that the $\Lambda$-dependent terms in
Eq.~(13) vanish in the $\Lambda\to\infty$ limit, thereby reproducing the
point hadron result. For finite cut-off, the first few terms in a small-$m^2$
expansion of the radii and magnetic moment are given by
$$
\eqalignno{
\rho^{\rm sachs}& = -{1\over 3}\Bigl({g\over 4\pi}\Bigr)^2(3-5\mbar^2)
\ln{m^2\over\Lambda^2} +\cdots\rightarrow \Bigl({g\over 4\pi}\Bigr)^2
\biggl[2-{1\over 3}(3-5\mbar^2)\ln{m^2\over\mns} +\cdots\biggr]&(14)\cr
\rho^{\rm dirac}& = -{1\over 3}\Bigl({g\over 4\pi}\Bigr)^2(3-8\mbar^2)
\ln{m^2\over\Lambda^2} +\cdots\rightarrow -{1\over 3}\Bigl({g\over
4\pi}\Bigr)^2
(3-8\mbar^2)\ln{m^2\over\mns} +\cdots &(15)\cr
\mu & = \Bigl({g\over 4\pi}\Bigr)^2\mbar^2
\ln{m^2\over\Lambda^2} +\cdots\rightarrow \Bigl({g\over 4\pi}\Bigr)^2
\biggl[-2+\mbar^2\ln{m^2\over\mns} +\cdots\biggr]\ \ \ , &(16)\cr}
$$
where $\mbar\equiv m/\mn$. Taking $m=\mpi$ and $g=\sqrt{2}\gpnn$ gives
the neutron EM charge radii and magnetic moment, while setting $m=m_\sst{K}$
and
$g=g_\sst{N\Lambda K}$ gives the strangeness radius and magnetic moment.
The expressions to the right of the arrows give the first few terms in a
small-$m^2$ expansion in the  $\Lambda\to\infty$ limit. The cancellation in
this limit
of the logarithmic dependence on $\Lambda$ arises from terms not shown
explicitly ($+\cdots$) in Eqs.~(14-16).

In the case of the axial form factor, we assume the pseudoscalar meson
loops to give the dominant correction to the bare isovector
axial matrix element of the nucleon. This assumption may be more justifiable
than in the case of the vector current form factors, since the lightest
pseudovector isoscalar meson which can couple to the nucleon is the $f_1$
with mass
1425 MeV. In this case one has $g_\sst{A}^{\rm phys}\approx g_\sst{A}^0[1+
\Delta_\sst{A}^{\pi}]$ and
$\eta_s= \coeff{3}{5}\Delta_\sst{A}^\sst{K}/ [1+\Delta_\sst{A}^{\pi}]$
where the $\coeff{3}{5}$ is just the spin-flavor factor relating
$f_\sst{A}^0$ and $g_\sst{A}^0$. The loop contributions are given
by the $\Delta_\sst{A}^{\pi ,\sst{K}}$, where
$$
\eqalignno{
\Delta_\sst{A}^{\sst{K} , \pi}& =
\pm\Bigl({g\over 4\pi}\Bigr)^2\Bigl[{\lambar^2\over 3}
\Bigl({\mbar^2-\lambar^2\over 4-\lambar^2}\Bigr)+{1\over 4}(2\mbar^2 -
\lambar^2)\ln \lambar^2 +\cdots\Bigr]&(17) \cr
&\longrightarrow {3\over 5}\Bigl({g\over 4\pi}\Bigr)^2\Bigl[-{1\over 2}\ln
\lambar^2+{5\over 4}+{1\over 2}\mbar^2+\cdots\Bigr]\ \ \ ,&(18) \cr}
$$
where $g=g_\sst{N\Lambda K}$ or $\gpnn$ as appropriate, and where $\lambar=
\Lambda/\mn$.
The upper (lower) sign corresponds to the kaon (pion) loop.
The opposite sign arises from the fact that $\Delta_\sst{A}^\pi$ receives
contributions from two loops, corresponding to a neutral and charged pion,
respectively. The isovector axial vector coupling to the intermediate
nucleon in these loops ($n$ and $p$) have opposite signs, while the
$\pi^\pm$-loop carries an additional factor of two due to the isospin
structure of the $NN\pi$ vertex.

For the scalar density, we obtain
$R_s = {\tilde\Delta^\sst{K}_\sst{S}}/[ 3 f_\sst{S}+
\Delta^\sst{K}_\sst{S}+\Delta^\pi_\sst{S}]$,
where the loop contributions are contained in
$$
\eqalignno{
\tilde\Delta^\sst{K}_\sst{S}& = \Bigl({g_\sst{N\Lambda K}\over 4\pi}\Bigr)^2
\Bigl[f_\sst{S} F_s^a(m_\sst{K}, \Lambda)+{\overline\gamma}_\sst{K}
F_s^b(m_\sst{K}, \Lambda)\Bigr] &(19)\cr
\Delta^\sst{K}_\sst{S} & = \Bigl({g_\sst{N\Lambda K}\over 4\pi}\Bigr)^2
\Bigl[3f_\sst{S}
F_s^a(m_\sst{K}, \Lambda)+2{\overline\gamma}_\sst{K}F_s^b(m_\sst{K}, \Lambda)
\Bigr]\ \ \ , &(20)\cr}
$$
and $\Delta^\pi_\sst{S}$, where the expression for the latter
is the same as that for
$\Delta^\sst{K}_\sst{S}$
but with the replacements $\bar\gamma_\sst{K}\to\bar\gamma_\pi$,
$m_\sst{K}\to\mpi$, and
$g_\sst{N\Lambda K}\to\sqrt{3}\gpnn$, and where
$\overline\gamma_{\pi , \sst{K}}=
\gamma_{\pi ,\sst{K}}/\mn$. The functions $F_s^a$ and $F_s^b$, which represent
the contributions from loops 1a (baryon insertion) and 1b (meson insertion),
respectively, are given by
$$
\eqalignno{
F_s^a(m, \Lambda)& = 2\Bigl({\lambar^2-\mbar^2\over 4-\lambar^2}\Bigr)
+{1\over 2}\mbar^2
\ln{m^2\over \Lambda^2}+\cdots\rightarrow \ln\lambar^2 -2 + {1\over 2}
\mbar^2\ln \mbar^2 +\cdots &(21)\cr
F_s^b(m, \Lambda) & = 1-{1\over
2}\Bigl({\lambar^2\mbar^2-\mbar^2-\lambar^2\over
\lambar^2-\mbar^2}\Bigr)\ln{m^2\over\Lambda^2}+\cdots\rightarrow
1+{1\over 2}(1-\mbar^2)\ln\mbar^2+\cdots\  . &(22)\cr}
$$

The expressions in Eqs.~(14-22) and curves in Fig. 2 lend themselves to
a number of observations. Considering first the vector and axial vector
form factors, we note that the mean-square radii display
significant SU(3)-breaking. The loop contributions to the radii
contain an I.R. divergence associated with the meson mass which manifests
itself as a leading chiral logarithm in Eqs.~(14-15). The effect is especially
pronounced for $\rho^{\rm dirac}$, where, for $\Lambda > 1$ GeV,
the results for $m=\mpi$ are roughly an order of magnitude larger
than the results for $m=m_\sst{K}$ (up to overall sign). In contrast, the
scale of SU(3)-breaking is less than a factor of three for the magnetic
moment and axial form factor over the same cut-off range. The chiral
logarithms which enter the latter quantities are suppressed by at least
one power of $\mbar^2$, thereby rendering these quantities
I.R. finite and reducing
the impact of SU(3)-breaking associated with the meson mass. Consequently,
in a world where nucleon strange-quark
form factors arose entirely from pseudoscalar meson
loops, one would see a much larger strangeness radius (commensurate with the
neutron EM charge radius) were the kaon as light as the pion than one would
see in the actual world. The scales of the strange magnetic moment and
axial form factor, on the other hand, would not be appreciably different
with a significantly lighter kaon.

	From a numerical standpoint, the aforementioned qualitative
features have some interesting implications for present and proposed
experiments. Taking the meson-nucleon form factor cut-off in the range
determined from fits to $BB'$ scattering, $1.2\leq\Lambda_{\rm Bonn}
\leq 1.4$ GeV, we find $\mustr$ has roughly the same scale as
the nucleon's isoscalar EM magnetic moment, $\mu^\sst{I=0}=\coeff{1}{2}
(\mu_p+\mu_n)$. The loop contribution is comparable in magnitude and
has the same sign as pole [14] and Skyrme [15] predictions. While the
extent to which the loop and pole contributions are independent and
ought to be added is open to debate, the scale of these two contributions,
as well as the Skyrme estimate, point to a magnitude for $\mustr$ that
ought to be observable in the SAMPLE experiment [6]. Similarly, the
loop and Skyrme estimates for $\eta_s$ agree in sign and rough order of
magnitude, the latter being about half the value extracted from the
$\nu p/\nubar p$ cross sections [3].
Under the identification of the strange-quark contribution
to the proton's spin $\Delta s$ with $\GAS(0)$, one finds a similar
experimental value for $\eta_s$ from the EMC data [4].

	In contrast, predictions for the strangeness
radius differ significantly between the models. In the case of the Sachs
radius, the signs of the loop and pole predictions differ. The sign
of loop predictions corresponds to one's naive expectation that the
kaon, having negative strangeness, exists further from the c.m. of the
$K-\Lambda$ system due to its lighter mass. Hence, one would expect
a positive value for $\rho^{\rm sachs}_s$ (recall that $\rho^{\rm sachs}$
and $<r^2>_{\rm sachs}$ have opposite signs). The magnitude of the
loop prediction for $\rho^{\rm sachs}_s$ is roughly $1/4$ to $1/3$ that
of the pole and Skyrme models and agrees in sign with the latter. In
the case of the Dirac radius, the loop contribution is an order of
magnitude smaller than either of the other estimates. From the
standpoint of measurement, we note that a low-$|q^2|$, forward-angle
measurement of the elastic $\evec p$ PV asymmetry, $\alr(\evec p)$,
is sensitive to the combination
$\rho_s^{\rm sachs}+\mustr=\rho^{\rm dirac}$ [28]. The asymmetry for
scattering from a $(J^\pi, I)=(0^+,0)$ nucleus such as $^4$He, on the
other hand, is
sensitive primarily to the  Sachs radius [28]. Thus, were the kaon
cloud to be the dominant contributor to the nucleon's vector current
strangeness matrix elements, one would not be able to observe them
with the $\alr(\evec p)$ measurements of Refs.~[7,8], whereas one potentially
could do so with the $\alr(^4\hbox{He})$ measurements of Refs.~[8,9].
Were the pole or Skyrme models reliable predictors of $\bra{N}\sbar
\gamma_\mu s\ket{N}$, on the other hand, the strangeness radii (Dirac
and/or Sachs) would contribute at an observable level to both types of
PV asymmetry. As we illustrate elsewhere [21], the
scale of the pole prediction is rather sensitive to one's assumptions
about the asymptotic behavior of the vector current form factors;
depending on one's choice of conditions, the pole contribution
could be significantly smaller in magnitude
than prediction of Ref.~[14]. Given these results, including
the difference in sign between the pole and both the loop and Skyrme
estimates, a combination of PV asymmetry measurements
on different targets could prove useful in determining which picture gives
the best description of nucleon's vector current strangeness content.

	From Eqs.(19-22), one has that the loop contributions to the
scalar density contain both U.V. and chiral singularities.
The U.V. divergence arises from the ${\bar q} q$ insertion in the intermediate
baryon line, while the chiral singularity appears in the loop containing
the scalar density matrix element of the intermediate meson. Despite the
chiral singularity, loop contributions to the matrix elements $m_q\bra{N}
{\bar q} q\ket{N}$ are finite in the chiral limit due to the
pre-multiplying factor of $m_q$. The ratio $R_s$ is also well-behaved
in this limit as well as in the limit of large $\Lambda$. For $\mpi\to 0$
and $m_\sst{K}\to 0$ simultaneously, one has $R_s \sim 1/8$, while for
$\Lambda\to\infty$, the ratio approaches $~1/12$. These limiting values
are independent of the couplings $f_\sst{S}$ and $\gamma_{\pi , \sst{K}}$
and are determined essentially by counting the number of logarithmic
singularities (U.V. or chiral) entering the numerator and denominator
of $R_s$ (note that we have not included $\eta$ loops in this analysis).
Consequently, the values for $R_s$ in the chiral and infinite cut-off
limits do not suffer from the theoretical ambiguities encountered in
the physical regime discussed in Section 2. It is also
interesting to observe that the limiting results have the same sign and
magnitude as the value of $R_s$ extracted from the $\Sigma$-term.

For $\Lambda\leq\Lambda_{\rm Bonn}$ and for $\mpi$ and $m_\sst{K}$ having
their physical values, the prediction for $R_s$ is smaller than in either
of the aforementioned limits and rather
dependent on the choice of $f_\sst{S}^0$ and $\gamma_\sst{M}$.
The sensitivity to the precise numerical values taken by these couplings
is magnified by a phase difference between $F_s^a$ and $F_s^b$. The
range of results associated with scenarios (A)-(C) is indicated in Table
I, with the largest values arising from choices (A) and (C). The change
in overall sign arises from the sign difference between loops 1a and 1b and the
increasing magnitude of $F_s^b$ relative to $F_s^a$
as $\Lambda$ becomes small. These results
are suggestive that loops may give an important contribution to the nucleon's
strange-quark scalar density, though the predictive power of the present
estimate is limited by the sensitivity to the input couplings. We emphasize,
however, that our estimates of the vector and axial vector form factors
do not manifest this degree of sensitivity.

We note in passing that scenario (C) gives a value for $g_\sst{A}^0
=g_\sst{A}[1+\Delta_\sst{A}^\pi(\Lambda)]^{-1}
\approx g_\sst{A}^0({\rm bag})$ for $\Lambda\sim\Lambda_{\rm Bonn}$.
The value obtained for $g_\sst{A}^0$ in this case depends only on
the assumption that pseudoscalar meson loops give the dominant correction
to $g_\sst{A}^0$ and involves no statements about the details of
quark model wavefunctions. In contrast, for $\Lambda\sim 1/R_{\rm bag}$
we find $g_\sst{A}^0\approx g_\sst{A}$. We also note that scenarios (A) and (C)
are consistent with the bag value for the scalar density renormalization
factor, $Z$. The latter is defined as $Z=\bra{H}\qbar q\ket{H}/N_q$, where
$N_q$ is the number of valence quarks in hadron $H$ [24, 27]. A test of
consistency, then, is the extent to which the equality $f_\sst{S}^0=Z$ is
satisfied. When $f_\sst{S}^0$ is computed using Eq.~(8), we find
$f_\sst{S}^0\approx 0.5$ in scenarios (A) and (C), while one has $Z_{\rm
bag}\approx 0.5$ [24]. These statements would seem to support the larger
values for $R_s$ in Table I.

As for the $\Lambda$-dependence of the form factors, we find that
the radii do not change significantly in magnitude over the range
$\Lambda_{\rm Bonn}\leq\Lambda\leq\infty$, owing  in part to the importance
of the chiral logarithm. The variation in the magnetic moment, whose
chiral logarithm is suppressed by a factor of $\mbar^2$, is somewhat
greater (about a factor of four for $m=m_\sst{K}$). The ratio $\eta_s$ is
finite as $\Lambda\to\infty$, with a value of $\approx -1$ in
this limit. This limit is approached only for $\Lambda >>$ the range of
values shown in Fig. 2d, so we do not indicate it on the graph.
The I.R. divergence ($\Lambda << 1\ \hbox{f}^{-1}$) in the
vector current quantities is understandable from Eq.~(13), which
has the structure of a generalized Pauli-Villars regulator. The impact
of the monopole meson-nucleon form factor is similar to that of including
additional loops for a meson of mass $\Lambda$. From the I.R. singularity
in the radii associated with the physical meson, one would expect a similar
divergence in $\Lambda$, but with opposite sign. The appearance of a
singularity having the same sign as the chiral singularity, as well as the
appearance of an I.R. divergence in the $\Lambda$-dependence of
magnetic moment which displays no chiral singularity, is due to the
derivative term in Eq.~(13). In light of
this strong $\Lambda$-dependence at very small values, as well as
our philosophy of taking as much input from independent sources
({\it viz}, $BB'$ scattering) we quote in Table 1 results for our loop
estimates using $\Lambda\sim\Lambda_{\rm Bonn}$.

It is amusing, nonetheless,
to compare our results for $m=m_\sst{K}$ and $\Lambda\sim 1\ \hbox{fm}^{-1}$
with those of the calculation of Ref.~[17], which effectively excludes
contributions from virtual kaons having wavelength smaller than the nucleon
size. Assuming this regime in the cut-off
is sufficiently far from the artificial zero at
$\Lambda=m_\sst{K}$ to be physically meaningful, our result for $\eta_s$
agrees in magnitude and sign with that of Ref.~[17].
In contrast, our estimates are about a factor of three larger for the
strangeness radii and a factor of seven larger for the strange magnetic moment.
We suspect that this disagreement is due, in
part, to the different treatment of gauge invariance in the two calculations.
In the case of the axial vector form factor, which receives no seagull
contribution, the two calculations agree. Had we omitted the seagull
contributions, our results for the Sachs radius would also have agreed.
For the Dirac radius our estimate would have been three times smaller and
for the magnetic moment three times larger than the corresponding estimates
of Ref.~[17]. At $\Lambda\sim\Lambda_{\rm Bonn}$, the
relative importance of the seagull for $\rho_s^{\rm sachs}$ and $\mustr$
is much smaller ($\sim$ 30\% effect) than at small $\Lambda$, whereas omission
of the seagull contribution to $\rho_s^{\rm dirac}$ would have reduced its
value by more than an order of magnitude. We conclude that the
extent to which one respects the requirements of gauge invariance at the
level of the WT Identity can significantly affect the results for loops
employing meson-nucleon form factors. We would argue that a calculation
which satsifies the WT Identity is likely to be more realistic that one
which doesn't and speculate, therefore, that the estimate of Ref.~[17]
represents an underestimate of
the loop contributions to $\rhostr$ and $\mustr$. An attempt to
perform a chiral-quark model calculation satisfying the WT Identity in order
to test this speculation seems warranted.

Finally, we make two caveats as to the limit of our calculation's
predictive power. First, we observe that for $\Lambda\sim\Lambda_{\rm Bonn}$,
the pion-loop gives a value for $\mu_n$ very close to the physical
value, but significantly over-estimates the neutron's EM charge
radii, especially $\rho_n^{\rm dirac}$ (see Fig. 2b). One would conclude,
then, that certainly in the case of mean-square radii, loops
involving only the lightest pseudoscalar mesons do not give a complete
account of low-energy nucleon form factor physics. Some combination
of additional loops involving heavier mesons and vector meson pole
contributions is likely to give a more realistic description of
the strangeness vector current matrix elements. In this respect, the
work of Ref.~[29] is suggestive.

Second, had we adopted the heavy-baryon chiral perturbation framework of
Ref.~[30], we would have kept only the leading non-analytic terms in $m$, since
in the heavy baryon expansion contributions of order $m^p$, $p=1,2,\ldots$
are ambiguous. Removal of this ambiguity would require computing order
$m^p$ contributions to a variety of processes in order to tie down the
coefficients of higher-dimension operators in a chiral lagrangian. The
present calculation, however, was not carried out within this framework
and gives, in effect, a model for the contributions analytic in $m$.
For $m=m_\sst{K}$, these terms contribute non non-negligibly to our
results. We reiterate that our aim is not so much to make reliable numerical
predictions as to provide insight into orders of magnitude, signs, and
qualitative features of nucleon strangeness. Were one interested in arriving at
more precise numerical statements, even the use of chiral perturbation
theory could be insufficient, since it appears from our results that pole
and heavier meson loops are likely to give important contributions to the
form factors. From this perspective, then,
it makes as much sense to include terms analytic in $m$ as to exclude them,
especially if we are to make contact with the previous calculations of
Refs.~[16, 17, and 20].

	In short, we view the present calculation as a baseline against
which to compare future, more extensive loop analyses. Based on a simple
physical picture of a kaon cloud, it offers a way, albeit provisional, of
understanding how strange quark matrix elements of the nucleon might
exist with observable magnitude, in spite of the success with which
constituent quark models of the nucleon account for its other low-energy
properties. At the same time, we have illustrated some of the qualitative
features of loop contributions, such as the impact of SU(3)-breaking in
the pseudoscalar meson octet, the sensitivity to one's form factor at
the hadronic vertex, and the importance of respecting gauge invariance at
the level of the WT Identity. Finally, when taken in tandem with the
calculations of Refs.~[14, 15], our results strengthen the rationale for
undertaking the significant experimental investment required to probe
nucleon strangeness with semi-leptonic scattering.
\medskip
\centerline{\bf Acknowledgements}
\medskip

It is a pleasure to thank S. J. Pollock, B. R. Holstein,
W. C. Haxton, N. Isgur, and E. Lomon for useful discussions.
\medskip
\centerline{\bf References}
\medskip
\item{1.}T.P. Cheng, {\it Phys. Rev. \bf D13} (1976) 2161.
\medskip
\item{2.}J. Gasser, H. Leutwyler, and M.E. Sainio, {\it Phys. Lett.
\bf B253} (1991) 252.
\medskip
\item{3.}L.A. Ahrens {\it et al.}, {\it Phys. Rev. \bf D35} (1987) 785.
\medskip
\item{4.}J. Ashman {\it et al.}, {\it Nucl. Phys. \bf B328} (1989) 1.
\medskip
\item{5.} D. B. Kaplan and A. Manohar, \NPB{310} (1988) 527.
\medskip
\item{6.}MIT-Bates proposal \# 89-06, Bob McKeown and D. H. Beck,
contact people.
\item{7.}CEBAF proposal \# PR-91-010, J.M. Finn and P.A. Souder,
	spokespersons.
\medskip
\item{8.}CEBAF proposal \# PR-91-017, D.H. Beck, spokesperson.
\medskip
\item{9.}CEBAF proposal \# PR-91-004, E.J. Beise, spokesperson.
\medskip
\item{10.}LAMPF Proposal \# 1173, W.C. Louis, contact person.
\medskip
\item{11.}S.J. Brodsky, P. Hoyer, C. Peterson, and N. Sakai, {\it Phys.
	Lett. \bf B93} (1980) 451; S.J. Brodsky, C. Peterson, and
	N. Sakai, {\it Phys. Rev. \bf D23} (1981) 2745.
\medskip
\item{12.}J. Collins, F. Wilczek, and A. Zee, {\it Phys. Rev. \bf
D18} (1978) 242.
\medskip
\item{13.}M. Burkardt and B.J. Warr, {\it Phys. Rev. \bf D45} (1992) 958.
\medskip
\item{14.}R. L. Jaffe, {\it Phys. Lett. \bf B229} (1989) 275.
\medskip
\item{15.}N. W. Park, J. Schechter and H. Weigel, {\it Phys. Rev.\/}
{\bf D43}, 869 (1991).
\item{16.}B.R. Holstein, in {\it Proceedings of the Caltech Workshop
on Parity Violation in Electron Scattering}, E.J. Beise and R.D.
McKeown, Eds., World Scientific (1990), pp. 27-43.
\medskip
\item{17.}W. Koepf, E.M. Henley, and S.J. Pollock, {\it Phys. Lett
\bf B288} (1992) 11.
\medskip
\item{18.} B. Holzenkamp, K. Holinde, and J. Speth, {\it Nucl. Phys.
\bf A500} (1989) 485.
\medskip
\item{19.}E. Hirsch {\it et al.}, {\it Phys. Lett. \bf B36} (1971) 139;
E. Hirsch, U. Karshon, and H.J. Lipkin, {\it Phys. Lett. \bf B36} (1971)
385.
\medskip
\item{20.}H.A. Bethe and F. deHoffman, {\it Mesons and Fields} (Row,
Peterson, and Co., Evanston, IL, 1955), Vol. II, p.289ff.
\medskip
\item{21.} M.J. Musolf and M. Burkardt, to be published.
\medskip
\item{22.}J.M. Gaillard and G. Sauvage, {\it Ann. Rev. Nuc. Part. Sci.
\bf 34} (1984) 351; D. Dubbers, W. Mampe, and J. D\"ohner, {\it Europhys.
Lett.} {\bf 11}
(1990) 195;S. Freedman, {\it Comments Nucl. Part. Phys.} {\bf 19} (1990)
209; M. Bourquin {\it et al.\/}, \ZPC{21} (1983) 27.
\medskip
\item{23.}A. Chodos, R.L. Jaffe, K. Johnson, and C.B. Thorn, {\it Phys.
Rev. \bf D10} (1974) 2599; T. DeGrand, R.L. Jaffe, K. Johnson, and
J. Kiskis, {\it Phys. Rev. \bf D12} (1975) 2060.
\medskip
\item{24.}J.F. Donoghue and K. Johnson, {\it Phys. Rev. \bf D21} (1980)
1975.
\medskip
\item{25.}J.F. Donoghue, E. Golowich, and B.R. Holstein, {\it Phys. Rep.
\bf 131} (1986) 319.
\medskip
\item{26.}R.G. Sachs, {\it Phys. Rev. \bf 126} (1962) 2256.
\medskip
\item{27.}S. Weinberg, in {\it Fetschrift for I. I. Rabi}, edited by
Lloyed Moltz (N. Y. Academy of Sciences, N.Y., 1977).
\medskip
\item{28.}M.J. Musolf and T.W. Donnelly, {\it Nucl. Phys. \bf A546}
(1992) 509.
\medskip
\item{29.} P. Geiger and N. Isgur, CEBAF Theory Pre-print \#CEBAF-TH-92-24
(1992).
\medskip
\item{30.}E. Jenkins and A.V. Manohar, {\it Phys. Lett. \bf B255} (1991)
558.
\medskip
\medskip
\centerline{\bf Captions}
\medskip
\noindent Fig. 1. Feynman diagrams for loop contributions to nucleon
strange quark matrix elements. Here, $\otimes$ denotes insertion of
the operator $\sbar\Gamma s$ where $\Gamma=1, \gamma_\mu$, or $\gamma_\mu
\gamma_5$. All four diagrams contribute to vector current matrix
elements. Only diagam 1a enters the axial vector matrix element.
Both 1a and 1b contribute to the scalar density.
\medskip
\noindent Fig. 2. Strange quark vector and axial vector parameters
as a function of nucleon-meson form factor mass, $\Lambda$. Here,
$\rhostr$ denotes the dimensionless Sachs (2a) and Dirac (2b) strangeness
radii. The strange magnetic moment is given in (2c). The axial vector
ratio $\eta_s$ is shown in (2d). Dashed curves indicate values of
these parameters for $\Lambda\to\infty$. The ranges corresponding
to the Bonn values for $\Lambda$ are indicated by the arrows.
The strong meson-nucleon
coupling $(g/4\pi)^2$ has been scaled out in (2a-c) and must
multiply the results in Fig. 2 to obtain the values in Table I.
\medskip
\vfill
\eject

\centerline{\bf Tables}
\medskip
$$\hbox{\vbox{\offinterlineskip
\def\strut{\hbox{\vrule height 15pt depth 10pt width 0pt}}
\hrule
\halign{
\strut\vrule#\tabskip 0.2cm&
\hfil$#$\hfil&
\vrule#&
\hfil$#$\hfil&
\vrule#&
\hfil$#$\hfil&
\vrule#&
\hfil$#$\hfil&
\vrule#&
\hfil$#$\hfil&
\vrule#\tabskip 0.0in\cr
& \multispan9{\hfil\bf TABLE I\hfil} & \cr\noalign{\hrule}
& \hbox{Source } && \rho_s^{\rm sachs} && \mustr && \eta_s && R_s
& \cr\noalign{\hrule}
& \hbox{elastic}\ \nu p/\nubar p\ [3]&& -&&- && -0.12\pm 0.07 && - & \cr
& \hbox{EMC}\ [4] && - && - && -0.154\pm 0.044 && - &\cr
& \Sigma_{\pi\sst{N}}\ [1,2] && - && - && - && 0.1\to 0.2& \cr
\noalign{\hrule}
& \hbox{kaon loops} &&0.41\to 0.49 &&-0.31\to -0.40 &&-0.029\to -0.041  &&
-0.007\to 0.047 &\cr
\noalign{\hrule}
& \hbox{poles}\ [14] && -2.12\pm 1.0&& -0.31\pm 0.009 && - && - &\cr
& \hbox{Skyrme (B)}\ [15] && 1.65 && -0.13 && -0.08 && - & \cr
& \hbox{Skyrme (S)}\ [15] && 3.21 && -0.33 && - && - & \cr
\noalign{\hrule}}}}$$
{\noindent\narrower {\bf Table I.} \quad Experimental determinations
and theoretical estimates of strange-quark matrix elements of the nucleon.
First two rows give experimental values for $\eta_s$,
where the EMC value is determined from the $s$-quark contribution to
the proton spin, $\Delta s$. Third row gives $R_s$ extracted from analyses of
$\Sigma_{\pi\sst{N}}$. Final four rows give theoretical estimates of
\lq\lq intrinsic' strangeness contributions in various hadronic models.
Loop values (row four) are those of the present calculation, where the
ranges correspond to varying hadronic form factor cut-off over the range
of Bonn values (see text). Final two rows give broken (B) and symmetric
(S) SU(3) Skyrme model predictions. First column gives dimensionless, mean
square Sachs strangeness radius. The dimensionless Dirac radius is
given by $\rho_s^{\rm dirac}=\rho_s^{\rm sachs}+\mustr$.\smallskip}

\vfill
\eject
\end{document}